\documentclass[showpacs,preprintnumbers,amsmath,amssymb,10pt]{ws-ijmpa}
\usepackage[super,compress]{cite}
\usepackage{graphicx}
\usepackage{epstopdf}
\usepackage{graphicx}
\usepackage{dcolumn}
\usepackage{bm}
\def\bea {\begin{eqnarray}}
\def\eea {\end{eqnarray}}

\def\be {\begin{equation}}
\def\ee {\end{equation}}

\begin{document}
\markboth{Golam Sarwar, Jan-e Alam}{Kinetic evolution and correlation of fluctuations in QGP }
\catchline{}{}{}{}{}
\title
{Kinetic evolution and correlation of fluctuations in an expanding quark gluon plasma}
\author{Golam Sarwar
}
\address{Variable Energy Cyclotron Centre-HBNI,\\ 1/AF, Bidhan Nagar, Kolkata - 700064, INDIA\\
golamsarwar@vecc.gov.in}
\author{Jan-e Alam}
\address{Variable Energy Cyclotron Centre-HBNI,\\ 1/AF, Bidhan Nagar, Kolkata - 700064, INDIA\\
jane@vecc.gov.in}
\maketitle
\begin{history}
\received{Day Month Year}
\revised{Day Month Year}
\end{history}
\begin{abstract}
Evolution of spatially anisotropic perturbation created in the system formed after Relativistic Heavy Ion Collisions 
has been studied. The microscopic evolution of the fluctuations 
has been examined within the ambit of Boltzmann Transport Equation (BTE) in a hydrodynamically 
expanding background.  The expansion of the background composed of Quark Gluon Plasma (QGP) is 
treated within the framework of relativistic hydrodynamics.
Spatial anisotropic fluctuations with  different geometry have been evolved through Boltzmann equation.
It is observed that the trace of such fluctuation survive the evolution.  Within the 
relaxation time approximation analytical results have been obtained for the 
evolution of  these anisotropies. Explicit relations between fluctuations  and transport 
coefficients have been derived.  The mixing of various Fourier (or $k$) modes of the 
perturbations during the evolution of the system has been explicitly demonstrated.
This study is very useful in understanding the  presumption that the measured anisotropies 
in the data from heavy ion collisions at relativistic energies imitate the initial state effects. 
The evolution of correlation function for the perturbation in pressure has been studied and shown that 
the initial correlation between two neighbouring points in real space evolves to a constant 
value at later time which gives rise to Dirac delta function for the correlation function
in Fourier space. The power spectrum of the fluctuation in thermodynamic quantities (like temperature 
estimated in this work) can be connected to the fluctuation in transverse momentum of the thermal 
hadrons measured experimentally. The bulk viscous coefficient of the QGP has been
estimated by using correlations of pressure fluctuation with the help of Green-Kubo relation.
Angular power spectrum of the anisotropies has been estimated in the appendix.

\keywords{QGP \and Fluctuation \and Kinetic Evolution \and Heavy Ion collision}
\end{abstract}

\ccode{PACS numbers: 25.75.-q,25.75.Dw,24.85.+p}

\section{Introduction}
The main aim of the heavy ion collision experiments at Relativistic 
Heavy Ion Collider (RHIC) and Large Hadron Collider (LHC) is to create a 
new state of matter called quark gluon plasma (QGP). Such a state of
matter, {\it i.e.} QGP \cite{pbm} may have existed in the early universe after a 
few microsecond of the Big Bang. One of the motivations to 
create and study  QGP in the laboratory is to understand the state of the
universe in the microsecond old era. The fluctuations of physical quantities
from their average values can be used to understand several properties
of the system {\it i.e.} the transport coefficients of the medium, the approach 
toward equilibrium, etc. The fluctuation becomes  extremely important in the neighborhood
of nuclear  phase transition.  The study of temperature ($T$) fluctuation in the 
cosmic microwave background radiation (CMBR)  has provided crucial 
information about the universe when it was about 
300,000 years old. This information has led to tremendous support to the Big Bang 
model of cosmology. The polarization of the photons resulting from the Thomson
scattering at the decoupling surface infected by density fluctuation gets reflected in 
the quadrupole moment of the phase space distribution of the incident photon. 
The  fluctuation of $T$ in the CMBR  is  introduced as a perturbation 
in the phase space distribution of photons. The evolution of this perturbation is studied
by using Boltzmann transport equation (BTE)~\cite{cmbr} in gravitational field 
with Thomson scattering in the collision term.  The linear polarization
resulted from the scattering is connected with the quadrupole
moment of the photon's phase space distribution.  

The fluctuations in the position of  nucleons (with finite size) in the 
colliding nuclei lead to lumpiness in the spatial distribution of
initial energy density of the system formed in Relativistic Heavy Ion Collisions Experiments (RHIC-E). The fluctuation in energy
density may also originate due to the energy deposition by the 
propagation of energetic partons produced in the early stage of the RHIC-E.
These fluctuations may lead to observable effects similar as temperature fluctuation in CMBR. 
We have adopted an approach similar to the one followed to study the fluctuation in CMBR.  
We study the evolution of perturbations 
by introducing a deviation, $\delta f(x,p,t)$ in the equilibrium 
distribution function, $f^{(0)}(p)$. The  bulk in equilibrium evolves via
relativistic hydrodynamics and the evolution of the fluctuations over the equilibrated expanding background,
$\delta f$ is treated within the ambit of BTE. In contrast to this the propagation
of perturbation has been studied using hydrodynamics in Refs.~\cite{pstaig,UW,kapusta,ripples,apm}.                                 
. $\delta f$ may be used to estimate
the fluctuations in various thermodynamical quantities as we will see below.
In the present work the equilibrated  background is assumed to be quark gluon plasma (QGP)
expected to be produced  in RHIC-E.  

The fluctuations in the thermodynamic quantities (e.g., hot spots created in the initial state of the collisions~\cite{qin})
can be related to perturbations in the phase space distributions in the hydrodynamic limit.  
The  evolution of these fluctuations can be analyzed within the ambit of BTE~\cite{degroot}. 
Fluctuations in thermodynamic quantities have been proposed as  signals
of the critical end point in the QCD phase transition~\cite{flc1,flc2}. Dissimilar 
fluctuations in partonic and hadronic phases in the net electric charge and baryon number
may shed light on the QCD phase transition in RHIC-E~\cite{asakawa}. Event by event fluctuations in the ratio of
positively to negatively charged pions may be used as an indicator of QCD transition~\cite{jeon1} as well as 
for understanding the  chemical equilibrium in the system formed in RHIC-E~\cite{jeon}. 
Evolution of these fluctuations near the critical end point has been studied by using BTE~\cite{flc3}. Kinetic theory approach has also been adopted to study fluctuations 
in particle and energy densities~\cite{flc4}. In Ref.~\cite{RJ} it has been argued that
the perturbations in hot QGP travel longer distance to reach the border of the medium
giving rise to the possibility of detectable signatures of these perturbations. 

The study of the fluctuations in the space time
structure of the fireball  driven by the fluctuations in the position of the 
nucleons in the colliding nuclei is an important contemporary issue in RHIC-E. 
Fluctuations in the space-time structure of the system will infect fluctuations
in the thermodynamic quantities. How  will  these fluctuations evolve with time in a hydrodynamically
expanding system and how are they connected with the transport coefficients for matter
formed in RHIC-E are addressed in this work. 

In Ref.~\cite{TrNeq} role of non-equlibrium processes on the evolution of QGP was studied
within the framework of Parton-Hadron String Dynamics (PHSD)  transport approach.
It was found that the event-by-event fluctuations on collective variables estimated by the
microscopic PHSD model is large due to non-equilibrium processes.   However, the ensemble averaged  results
from these events is close to the results obtained in (2+1)-dimensional viscous hydrodynamics.  
In this context the study of the evolution  of perturbations  in the
microscopic approach is crucial. Therefore, the present study of the perturbations in a hydrodynamically expanding background 
within the ambit of kinetic theory is appropriate. This is  not only expected to achieve better microscopic understanding
 of the physics  but also avoid question of breaking down of hydrodynamic description of fluctuation~\cite{HdrBrk}.
Authors in Ref.~\cite{JtEdp} has  discussed deposition of energy by the away side mini-jet in the non-equilibrium framework.
Corresponding perturbation in the medium has clear anisotropic form in space, as mini-jet deposits energy along its path.
Therefore,  instead of Gaussian type perturbation  consideration of anisotropic perturbation in space will be more appropriate in such
cases. In this work we have developed a formalism to find the evolution of fluctuation in hydrodynamically expanding background
such that microscopically mode-by-mode analysis can be performed.

The paper is organized as follows. In the next section we  discuss the evolution of fluctuations in 
a non-expanding background and present a relation between energy fluctuation in Fourier space 
and viscous coefficient. Section 3 is devoted to discuss the progression of fluctuations in a hydrodynamically
expanding QGP background.  Results are presented in section 4 and section 5 is dedicated to  summary and discussions.
\section{Evolution of fluctuations in a non-expanding background}
In the following subsections we discuss the connection of $\delta f$, a small deviation of phase space 
distribution from its equilibrium value with the fluctuations in various 
thermodynamic variables and its time evolution in a non-expanding background within the framework of BTE (results with 
expansion will be discussed in section 3).
The phase space distribution function, $f(\vec{x},\vec{p},t)$
of a system slightly away from equilibrium, at time $t$, position $\vec{x}$,
momentum $\vec{p}$ can be written as~\cite{eml},
\begin{equation}
f(\vec{x},\vec{p},t )=f^{(0)}(p)\{1+\Psi(\vec{x},\vec{p},t)\}=f^{(0)}(p)+f^{(0)}\Psi(\vec{x},\vec{p},t)
\label{eq1}
\end{equation}   
where  $f^{(0)}(p)$ is the phase space distribution function  in equilibrium and
$\Psi(\vec{x},\vec{p},t)$ is the 
fractional deviation from $f^{0}(p)$. 
$\Psi(\vec{x},\vec{p},t)$ can be used to estimate the fluctuations in various
thermodynamic quantities in the system.  
The evolution of $\Psi$ is  governed by BTE,
which in turn provides the  relation between 
the dissipative effects and fluctuations in the hydrodynamic limit. 

\subsection{Fluctuations of various hydrodynamic quantities in Fourier space}
A given fluctuations in spatial coordinate can be expressed in terms of various $k$-modes
in Fourier space. These $k$-modes or wave number modes can be connected to the
wave length ($\lambda=2\pi/k$) modes, which in turn is related to the angular size ($\vartheta$) of
the fluctuations through the relation: $\vartheta=\lambda/d$, where $d$
is the angular diameter as is usually done in analyzing temperature fluctuations in the 
universe.  Fourier analysis is also important to understand what are the $k$-modes of the 
fluctuations that dissipate during the evolution of the system. For example it is important
to determine the viscous horizon in heavy ion collisions~\cite{pstaig}. 

The energy momentum tensor, $T^{\mu\nu}$ of the system under study can be written as:
$T^{\mu\nu}=\overline{T}^{\mu\nu}+\Delta T^{\mu\nu}$,
where the equilibrium (ideal) part, $\overline{T}^{\mu\nu}$ is 
determined by $f^0(p)$ and the dissipative part,
$\Delta T^{\mu\nu}$ is determined by $\Psi$, {\it i.e.} 
\begin{equation}
T^{\mu\nu}=\int d^3p\,\frac{p^{\mu}p^{\nu}}{p^0}f(\vec{x},\vec{p},t)
\label{eq2}
\end{equation}
where $f(\vec{x},\vec{p},t)$ is given by Eq.~\ref{eq1}.
The $\overline{T}^{\mu\nu}$ of the system in equilibrium can be obtained
from Eq.~\ref{eq2} by setting $\Psi=0$. 
The metric, $g^{\mu\nu}$, in Minkowski space-time is taken as $g^{\mu\nu}=(-1,1,1,1)$.
We assume that the momentum, $\vec{p}$ can be written as $p_i=pn_i$, 
{\it i.e.} $\vec{p}=|\vec{p}|\hat{n}$ where $\hat{n}$ is a unit vector and
$d^3p=p^2dp \,d\Omega$, $d\Omega$ being solid angle associated with 
$n_i$ which satisfies $\int d\Omega n_in_j=4\pi{\delta}_{ij}/3 $ and $\int d\Omega n_in_jn_k=0$. 

It is straightforward to obtain various components of $T^\mu_\nu$ from Eq.~\ref{eq2}.  
The deviation of the components of the stress energy tensor, $\Delta T^{\mu\nu}$ from their ideal 
values can be expressed in  terms of the perturbation, $\Psi$  as follows:
\begin{equation}
\begin{aligned}
\;\;\;\Delta T^{0}_{0} &= -\int p^2dp \,d\Omega\,\epsilon\,f^{(0)}(p)\Psi(\vec{x},\vec{p},t),\\
\Delta T^{0}_{i} &= \int p^2dp \,d\Omega\,pn_if^{(0)}(p)\Psi(\vec{x},\vec{p},t),\\
\Delta T^{i}_{j} &=  \int p^2dp \,d\Omega\,\frac{p^2}{\epsilon}\,n^in_j\,f^{(0)}(p)\Psi(\vec{x},\vec{p},t)
\end{aligned}
\label{eq3}
\end{equation}
where  $\epsilon=p^0=-p_0=\sqrt{p^2+m^2}$ and  $m$ is the mass of the particle. 
It is to be noted that the  integral, $ \int d^3p\,pn_if^{(0)}(p)=0$, since in equilibrium 
the system is isotropic 
{\it i.e.} all directions are equally probable in its rest frame, the  phase space average of 
momentum vector then is zero. 
 
The ideal part of the stress energy tensor in the hydrodynamic limit is given by, 
\begin{equation}
\overline{T}^{\mu}_{\nu}=(\bar{\rho}+\bar{P})U^{\mu}U_{\nu}-\bar{P}g^{\mu}_{\nu},
\label{eq4}
\end{equation}
where $U^{\mu}={dx^\mu}/{d\tau}=\gamma(1,\vec{v})$ is the four velocity of the fluid and  $\tau$ is the proper time.  
The relations among various components of $\overline{T}^{\mu\nu}$ and the thermodynamic
variables are:
$\overline{T}^{0}_{0} =-\bar{\rho},\,\, \overline{T}^{0}_{i} =-\bar{T}^{i}_{0}=0,\,\,$
and $\overline{T}^{i}_{j} =\bar{P}\delta^{i}_{j}$.
Therefore,
$\bar{\rho}=\int p^2dp\,d\Omega\,\epsilon f^{(0)}(p)$, is the average energy density. 
Other thermodynamic quantities like pressure, number density etc can be estimated in a similar way. 
Equilibrium distribution is isotropic, therefore,  integration over $d\Omega$ will simply give $4\pi$.
If the fluid is slightly away from equilibrium  with space time dependent 
fluctuations in  energy density,  pressure, velocity, etc., then the system will evolve toward equilibrium 
through dissipative processes. 
In such situation the components of the energy momentum tensors can be  explicitly expressed in terms of the 
thermodynamic variables as:
\begin{equation}
\begin{aligned}
{T}^{0}_{0}(x_i,t) &=-\{\bar{\rho}+ \delta \rho(x_i,t)\},\\ 
{T}^{0}_{i}(x_i,t) &=-{T}^{i}_{0}=(\bar{\rho}+\bar{P})v_i,\\
{T}^{i}_{j}(x_i,t)  &=\{\bar{P}+\delta P(x_i,t) \}\delta^{i}_{j}+\Sigma^i_j(x_i,t),\\  
\Delta T^i_i=0,  
\end{aligned}
\label{eq5}
\end{equation}
where $v_i$ is the $i^{th}$ component of the velocity perturbation.
We can choose a frame which is moving with velocity close to the  velocity of the fluid,
 so that the fluid velocity measured from this frame is small.
From Eq.~\ref{eq5}, we get $\Sigma^i_j= {T}^{i}_{j}-\delta ^{i}_{j}{T}^{k}_{k}/3$.
By using Eqs.~\ref{eq3}, ~\ref{eq4} and ~\ref{eq5} we get, 
\begin{equation}
\begin{aligned}
   \delta \rho(x_i,t) &=-\delta {T}^{0}_{0}(x_i,t), \\
   v_i(x_i,t)&=\frac{\delta {T}^{0}_{i}(x_i,t)}{(\bar{\rho}+\bar{P})},\\ 
    \delta {T}^{i}_{j}(x_i,t)&=\delta P(x_i,t) \delta^{i}_{j}+\Sigma^i_j(x_i,t),
 \end{aligned}
 \label{eq6}
\end{equation}
with $\delta P(x_i,t)=\delta {T}^{i}_{i}(x_i,t)/3$.  
The shear stress,
$\Sigma^i_j(x_i,t)$ can be expressed in terms of shear viscous coefficient, $\eta$ as
$\Sigma^i_j(x_i,t)=-\eta(\frac{\partial U_i}{\partial x_j}
+\frac{\partial U_j}{\partial x_i}-\frac{2}{3}\delta^i_j\frac{\partial U_l}{\partial x_l})$ and 
the thermal conductivity ($\chi$) is defined through the relation, 
$\delta T^0_j=-\chi (\frac{\partial T}{\partial x_i}+T\frac{\partial U}{\partial t})$ ~\cite{weinberg}.

It is useful to express these quantities, {\it i.e.} various components of $\delta T^{\mu\nu}$ 
in Fourier or  $k$-space because expansion of these quantities in terms the spherical harmonics ($Y_{lm}$)
will enable to connect the angular scales set by $l$ in terms of $k$ analogous to the determination of angular scale in CMBR
\cite{cmbr}. 
In $k$-space these quantities are  marked by tilde $(\,\tilde{}\,)$ as: 
\begin{equation}
\tilde{\Sigma}^l_j(k_i,t)=-i\eta( \tilde{U}^lk_j+ 
\tilde{U}_jk^l-\frac{2}{3}\delta^l_jk^r\tilde{U}_r)
\end{equation}
and
\begin{equation}
\delta\tilde{T}^0_i(k_i,t)=-\chi\left[ik_i\tilde{T}(k_l,t)+\tilde{T}(k_l,t)\frac{\partial \tilde{U}_i}
{\partial t}\right].
\end{equation}

By using  Eqs.~\ref{eq3} and \ref{eq6} the fluctuations in $k$-space can be expressed in terms of
Fourier mode, $\Psi$ as: 
\begin{equation}
\begin{aligned}
\delta\tilde{\rho}(k_i,t)&= \int p^2dp \,d\Omega\,\epsilon \,f^{(0)}(p)\tilde{\Psi}(k_i,p,n_i,t),\\
               \tilde{v}_i(k_l,t)&=-\frac{1}{(\bar{\rho}+\bar{P})} \int p^2dp \,d\Omega\,pn_if^{(0)}(p)\tilde{\Psi},\\
               \tilde{\Sigma}^i_j(k_l,t)&=\int p^2dp \,d\Omega\,\frac{p^2}{\epsilon}(n_in_j-\frac{1}{3}\delta_{ij})
f^{(0)}(p)\tilde{\Psi},\\
               \delta\tilde{P}(k_i,t)&=\frac{1}{3}\int p^2dp \,d\Omega\,\frac{p^2}{\epsilon}f^{(0)}(p)\tilde{\Psi}
\end{aligned}
\label{eq7}
\end{equation}
where $\tilde{\Psi}$ is the Fourier transform of $\Psi$. Now we take the  zenith direction along $\vec{k}$ 
and then the angular dependence of $\tilde{\Psi}(k_i,p,n_i,t)$ can be expressed in terms of 
angles between $\hat{k}$ and $\vec{n}$.  
Depending on the symmetries of the problem under consideration  
$\tilde{\Psi}$ can be expanded  in a series of 
suitably chosen basis  functions e.g, for axial symmetry in terms of  
Legendre polynomials and in absence of  such symmetry it can be 
expressed in terms of  spherical harmonics~\cite{itoh}.

The vector component $n_i$ and tensor components $(n_in_j-\frac{1}{3}\delta_{ij})$,
appearing in the expressions  for $v_i(k_l,t)$ and $\Sigma^i_j(k_l,t)$ respectively,
can be converted into functions of $\theta$ (angle between $\vec{k}$ and $\vec{n}$), 
by taking contraction with suitable tensors made out of the components of $\hat{k}$. 
If we contract $n_i$ with $k_i$ then we get $k\hat{k}\cdot\hat{n}=k \cos\theta=kP_1(\hat{k}\cdot\vec{n})$ 
and by contracting $(n_in_j-\frac{1}{3}\delta_{ij})$ with $(\hat{k_i}\hat{k_j}-\frac{1}{3}\delta_{ij})$ 
we get $\frac{2}{3}\frac{1}{2}(3(\hat{k}\cdot
\hat{n})^2-1)=\frac{2}{3}\frac{1}{2}(3\cos^2\theta-1)=\frac{2}{3}P_2(\hat{k}\cdot\hat{n})$, 
where $P_l$ s are Legendre polynomials. For axial symmetric distribution of $\vec{p}$ it helps to
connect different co-efficient of expansion of $\tilde{\Psi}$ in terms of Legendre polynomials with corresponding 
scalar quantities obtained from  $v_i(k_l,t)$ and $\Sigma^i_j(k_l,t)$ due to 
orthogonality relation satisfied by $P_l$'s.  
We define the scalar  quantities like $\Delta$, $\theta$ and $\sigma$ as~\cite{cpma}, which, as will be seen later
allow us to get evolution equation for the Fourier modes.  
The fluctuation in energy density in Fourier space is given by, 
\begin{equation}
\Delta(k_i,t)=\frac{\delta{\tilde{\rho}}(k_i,t)}{\bar{{\rho}}}=
-\frac{\delta{\tilde{T}}^0_0(k_i,t)}{\bar{{\rho}}},
\label{eq6a}
\end{equation}
Similarly we define energy flux,
\begin{equation}
\theta(k_i,t)=ik^j\tilde{v}_j=\frac{ik^j\delta{\tilde{T}}^{0}_{j}(k_i,t)}{({\bar{\rho}}+{\bar{P}})},
\label{eq7a}
\end{equation}
and the shear stress as,
\begin{equation}
({\bar{\rho}}+{\bar{P}})\sigma(k_l,t)=-(\hat{k_i}\hat{k_j}-\frac{1}{3}\delta_{ij})
\tilde{\Sigma}^i_j(k_l,t),
\label{eq8a}
\end{equation}
The quantity, $\theta(k_i,t)=ik^jv_j$ originates from the velocity gradient.
$\theta$ and $\sigma$ can be expressed in terms of the shear viscous coefficient ($\eta$) and thermal 
conductivity ($\chi$) as follows:
\begin{equation}
\begin{aligned}
\sigma(\vec{k},t)&=-\frac{4}{3}\frac{\eta}{{\bar{\rho}}+{\bar{P}}}\,ik^j\tilde{v}_j(\vec{k},t)\\
\end{aligned}
\end{equation}
and 
\begin{equation}
\begin{aligned}
\theta(\vec{k},t) &=\frac{\chi}{{\bar{\rho}}+{\bar{P}}}(k^2 \tilde{T}(k_l,t)
-ik_l\tilde{T}(k_l,t)\dot{\tilde{v}}_l(k_l,t)).  
\end{aligned}
\label{eq9}
\end{equation}
The left hand side of both the equation above can be estimated from the solution BTE. 
Fourier transformation of these equations in frequency space will lead to dispersion
relation.  This relation can be used to determine  those $k$ (wave number) values  which will 
dissipate  due to viscous effects. 

Now Eqs.~\ref{eq7}, \ref{eq6a}, \ref{eq7a} and \ref{eq8a} can be used to obtain the fluctuations in the 
energy density, pressure and velocity in $k$-space as: 
\begin{equation}
\begin{aligned}
                   \Delta (k_i,t)
                 &=\frac{1}{4\pi}\int \,d\Omega\, 
\frac{\int p^2dp \,\epsilon f^{(0)}(p)\tilde{\Psi}(k_i,p,n_i,t)}{\int p^2dp\,\epsilon f^{(0)}(p)},\\
                   \frac{\delta\tilde{P}(k_i,t)}{{\bar{P}}} &= \frac{1}{4\pi}\int \,d\Omega\, 
\frac{\int p^2dp \,(p^2/\epsilon) 
f^{(0)}(p)\tilde{\Psi}(k_i,p,n_i,t)}{\int p^2dp\,(p^2/\epsilon) f^{(0)}(p)},\\
                \theta(k_i,t)&=\frac{ik}{4\pi}\int \,d\Omega \, (\hat{k}.\hat{n})\,\frac{\int p^2dp\,
f^{(0)}(p)\tilde{\Psi}(k_i,p,n_i,t)}{\int p^2dp\,(\epsilon +p^2/3\epsilon)f^{(0)}(p)},\\
              \sigma(k_i,t)&=-\frac{1}{4\pi}\int d\Omega \,((\hat{k}.\hat{n})^2-\frac{1}{3})
\frac{\int p^2dp \, f^{(0)}(p)\tilde{\Psi}(k_i,p,n_i,t)}{\int p^2dp\,(\epsilon +p^2/3\epsilon)f^{(0)}(p)}.\\
\end{aligned}
\label{eq10}
\end{equation}
To understand the angular scale determined by the multipole number $l$ (as used
in the appendix to find the angular correlations) 
we expand $\tilde{\Psi}$ in terms of Legendre polynomials for
an axially  symmetric distribution as:
\begin{equation}
\tilde{\Psi}(\vec{k},\hat{n},p,t)=\sum_{l=0}^{\infty}(-i)^l(2l+1)\Psi_l(\vec{k},p,t)P_l(\hat{k}.\hat{n}),
\label{eq11}
\end{equation}
where the factor$ (-i)^l(2l+1)$ is used to simplify the expansion of a plane wave form of $\tilde{\Psi}$. 
The $l$ is related to angular resolution of the anisotropies, {\it i.e.} smaller angular scale will require
larger $l$ and vice versa.  The temperature fluctuations ($\Delta T$) may be obtained from Eq.~\ref{eq11} by using the
relation~\cite{cpma}:
\begin{equation}
\Delta T/\bar{T}=-(\partial lnf^{(0)}/\partial lnp)^{-1}\Psi
\label{delT}
\end{equation}
For simplicity we will consider the massless limit, $m=0$ which gives the relation $\epsilon=p$.
The energy density is a scalar quantity  whereas the velocity 
and the shear tensor are vector and tensor respectively and these aspects are also
bound to reflect in  the corresponding fluctuations. Therefore, the orthogonality of 
$P_l$'s ensure that the fluctuations in scalar, vector and tensor  quantities are dictated by the coefficients
$\Psi_0$, $\Psi_1$ and $\Psi_2$ which are  obtained by 
substituting $\Psi$  from Eq.~\ref{eq11} in Eq.~\ref{eq10} and performing the angular integration as,
\begin{equation}
\begin{aligned}
\Delta(k_i,t)&={\delta\rho(k_i,t)}/{\bar{\rho}}=\int p^2dp \,p f^{(0)}(p)\Psi_0(k_i,p,t)/\bar{\rho},\\
&{\ \ \  }{\delta P(k_i,t)}/{\bar{P}}=\int p^2dp \,p f^{(0)}(p)\Psi_0(k_i,p,t)/\bar{\rho}\\
\theta(k_i,t)&={3ik^j\delta {T}^{0}_{j}}/{(4\bar{\rho})}=\frac{3}{4}k\int p^2dp \,p 
f^{(0)}(p)\Psi_1(k_i,p,t)/\bar{\rho}\\
\sigma(k_l,t)&=\frac{1}{2}\int p^2dp \,p f^{(0)}(p)\Psi_2(k_i,p,t)/\bar{\rho}
\end{aligned}
\label{eq13}
\end{equation}
where $\bar{\rho}=\int p^2dp \,p f^{(0)}(p)$.
The above set of equations  can be written in a more compact form through the expansion of
the function $F(\vec{k},\hat{n},t)$ which is obtained by integrating $\delta f$ over the magnitude of
momentum, $\vec{p}$.
\begin{equation}
F(\vec{k},\hat{n},t)=\int p^2dp \,p f^{(0)}(p)\Psi(\vec{k},p,n_i,t)/\bar{\rho}
\label{eq14}
\end{equation}
therefore, $F$ has the angular dependence of $\Psi$ and consequently $F$ can be expressed as:
\begin{equation}
F(\vec{k},\hat{n},t)=\sum_{l=0}^{\infty}(-i)^l(2l+1)F_l(\vec{k},t)P_l(\hat{k}.\hat{n}).
\label{eq15}
\end{equation}
with
\begin{equation}
F_l(\vec{k},t)=\int p^2dp \,p f^{(0)}(p)\Psi_l(\vec{k},p,t)/\epsilon_0
\label{eq16}
\end{equation}
The fluctuations in terms of $F_l$'s are now given by
\begin{equation}
\Delta(\vec{k},t)= F_0(\vec{k},t),\,\, {\delta P(\vec{k},t)}/{\bar{P}}= F_0(\vec{k},t),\,\, \theta(\vec{k},t)={3}/{4}kF_1(\vec{k},t),\,\,
\sigma(\vec{k},t)={1}/{2}F_2(\vec{k},t).
\label{eq17}
\end{equation}
Using the relation $\sigma(\vec{k},t)=-{4\eta ik^jv_j(\vec{k},t)}/{3(\bar{\rho}+\bar{P})}$, 
and writing $ ik^jv_j(\vec{k},t)=\Theta (\vec{k},t)$ we get
an important relation which connects 
the fluctuation ($F_2$) with the transport coefficient ($\eta$), 
\begin{equation} 
F_2(\vec{k},t)=-\frac{8\eta}{3(\bar{\rho}+\bar{P})}\Theta (\vec{k},t)\equiv -\frac{8}{3\bar{T}}
\frac{\eta}{\bar{s}}\Theta (\vec{k},t) 
\label{f2}
\end{equation} 
where the thermodynamic relation, $\bar{h}=\bar{\rho}+\bar{P}=\bar{s}\bar{T}$, 
among enthalpy density ($\bar{h}$), entropy density ($\bar{s}$) and  temperature 
(${\bar{T}}$)  has been used. 
 The $\eta$ appears as a coefficient of 2$^{nd}$ rank tensor
involving gradient in the $i$th direction of the $j$th component of velocity, as a result the $l=2$ term
appears in the expression for $\eta$ in Eq.~\ref{f2}. In a similar way, using Eq.~\ref{eq17} 
the bulk viscosity $\zeta$ can be related to the fluctuation in pressure ($F_0$) as:
$\delta\,P=-ik_l v^l\,\zeta$.

\subsection{Fluctuations in Fourier space and transport coefficients in relaxation time approximation}

The temperature fluctuations, $\Delta T(\theta,\phi)$ in CMBR is generally expanded in 
Laplace series in terms of spherical harmonics, $Y_{lm}(\theta,\phi)$.  The maximum value of $l$ is 
determined by the angular resolution of the detector which can be connected to the wave number 
($k$) corresponding to the Fourier transform of the spatial anisotropy. Therefore, in analogy with
fluctuations in the CMBR the spatial anisotropy is studied here in Fourier space.
But first we briefly discuss it in coordinate space.

The BTE, $p^\mu\partial_\mu f= C[f]$ in absence of external force 
and in the relaxation time approximation (similar approximation were used {\it e.g.} in Refs~\cite{FRS1,FRS2,Hatta})  reduces to 
\begin{equation}
\frac{\partial \Psi }{\partial t}+\frac{p^i }{\epsilon}\frac{\partial \Psi}{\partial x^i}=-\frac{\Psi(\vec{x},\vec{p},t )}
{\tau_R}.
\label{eq20}
\end{equation}
for $\Psi$.
In Eq. \ref{eq20} $\tau_R$ is the relaxation time. In this work we assume
that the system is close to the local equilibrium and the 
collisions between the particles bring the system back to
the equilibrium within a time  scale $\tau_R$. For the present work
the relaxation time can be estimated as the inverse of the reaction rate
of the quarks and gluons using pQCD cross sections and Hard Thermal Loop 
Approximations~\cite{thoma}. 
The solution of Eq. ~\ref{eq20} for a given initial (at time $t_0$) 
distribution, $\Psi_{in}(\vec{x}, \vec{p},t_0)$ is~\cite{hlevine}:
\begin{equation}
\Psi(\vec{x},\vec{p},t-t_0)=\Psi_{in}\left((\vec{x}-\frac{\vec{p}}{p_0}(t-t_0)),\vec{p}\right)\exp \left[-\frac{(t-t_0)}{\tau_R}\right]
\label{eq22a}
\end{equation}
Knowing $\Psi$ it is straightforward to estimate the fluctuation in energy density from the following expression:
\begin{equation}
\Delta(\vec{x},t-t_0)=\frac{\int p^2dp \,d\Omega\,\epsilon f^{(0)}(p)\Psi(\vec{x},\vec{p},t-t_0 )}{\int p^2dp \,d\Omega\,\epsilon f^{(0)}(p)}.
\label{fluc1}
\end{equation}
The  solution of Eq.~\ref{eq20} given by Eq.~\ref{eq22a} is useful to study the time evolution of spatial anisotropy of the matter.

Now we would like to derive a  relation between the  
fluctuation in energy density  and transport coefficients. To facilitate this 
we write Eq.~\ref{eq20} for massless particles
(as the case may be for partonic plasma produced in RHIC-E) in $k$-space:
\begin{equation}
\frac{\partial \Psi }{\partial t}+ik(\hat{k}.\hat{n}){ \Psi }=-\frac{\Psi(\vec{k},\vec{p},t )}{\tau}.
\label{eq21}
\end{equation}
With the help of Eq.~\ref{eq14}, Eq.~\ref{eq21} can be  reduced to an equation describing the time evolution of $F$ as
\begin{equation}
\frac{\partial F}{\partial t}+ik(\hat{k}.\hat{n}){ F}=-\frac{F(\vec{k},t )}{\tau}.
\label{eq22}
\end{equation}
This equation has the following solution, 
\begin{equation}
F(\vec{k},\hat{n},t)=F(\vec{k},\hat{n},t_0)\exp\left[-(\frac{1}{\tau}+ik\mu)(t-t_0)\right]
\label{eq25}
\end{equation}
where $\hat{k}.\hat{n}=\mu$. The value of $F(\vec{k},\hat{n},t)$ can be obtained from its value 
at initial time, $t_0$. Eq.~\ref{eq25} is a general expression for the fluctuations 
in the sense that all the quantities, {\it e.g.} $\Delta$, $\theta$, $\sigma$, 
discussed above at time $t$ can be obtained from this expression if their corresponding 
initial values are supplied.  
Expanding $F(\vec{k},\hat{n},t)$ as in  Eq.~\ref{eq15} and using the orthogonality relations of 
$P_l(\mu)$s we get,
\begin{equation}
F_l(\vec{k},t)=\frac{1}{2}e^{-\frac{(t-t_0)}{\tau}}\sum_{s=0}^{\infty}(-i)^{(s-l)}(2s+1)F_s(\vec{k},t_0)
\int_{-1}^{+1}d\mu P_l(\mu) P_s(\mu)e^{-ik\mu (t-t_0)}.
\label{eq26}
\end{equation}

For $l=0,1,2$ we have, 
\[
  \begin{pmatrix} F_0(k,t) \\  F_1(k,t)\\ F_2(k,t) \end{pmatrix}= \begin{pmatrix} I_0 & (-i)3 I_1 &(-5)\frac{1}{2}(3I_2-I_0) \\  I_1 & (-i)3 I_2 &(-5)\frac{1}{2}(3I_3-I_1) \\ \frac{1}{2}(3I_2-I_0)  & (-i)\frac{1}{2}(I_3-I_1)&(-5)\frac{1}{4}(9I_4-6I_2+I_0) \end{pmatrix} 
\begin{pmatrix}  F_0(k,t_0) \\  F_1(k,t_0)\\ F_2(k,t_0), \end{pmatrix}
\]
where $ I_n \equiv I_n(k,t)$ and $\alpha=k\,(t-t_0)$
$$I_n(\alpha)=\int_{-1}^{+1} d\mu \, \mu^n e^{-i\mu\alpha },$$ 
for $n=0$ $I_0$ is given by
$$I_0(\alpha)=(-i)\,\, 2\,\, \frac{\sin\alpha}{\alpha}$$
The following relations may be used to obtain $I_j(\alpha)$ for $j=1,2,....$
$$I_{n+k}(\alpha)=\frac{1}{(-i)^k} \frac{d^k I_n}{d^2 \alpha}.$$

Therefore, the energy density fluctuation at time $t$ is given by:
\begin{equation}
\Delta(\vec{k},t)=\frac{1}{2}e^{-\frac{(t-t_0)}{\tau}}\sum_{s=0}^{\infty}(-i)^{s}(2s+1)F_s(\vec{k},t_0)
\int_{-1}^{+1}d\mu P_s(\mu)e^{-ik\mu (t-t_0)}.
\label{eq27}
\end{equation}
Taking terms upto $s=2$ in the expression for $\Delta(\vec{k},t)$ we get, 
\begin{equation}
\Delta(\vec{k},t)=\frac{1}{2}e^{-\frac{(t-t_0)}{\tau}}\sum_{s=0}^{2}(-i)^{s}(2s+1)F_s(\vec{k},t_0)
\int_{-1}^{+1}d\mu P_s(\mu)e^{-ik\mu (t-t_0)}.
\label{eq28}
\end{equation}
Performing the integration over $\mu$ we get,
\begin{equation}
\begin{aligned}
\Delta (\vec{k},t)=&e^{-\frac{(t-t_0)}{\tau}}[ F_0(\vec{k},t_0) \{ \frac{\sin k(t-t_0)}{k(t-t_0)}\}
                                 + 3F_1(\vec{k},t_0)\{\frac{\cos k(t-t_0)}{k(t-t_0)}- \frac{\sin k(t-t_0)}{\{k(t-t_0)\}^2}\}\\
                              &  -5F_2(\vec{k},t_0 )
                              \{  \frac{\sin k(t-t_0)}{k(t-t_0)}+\frac{3\cos k(t-t_0)}{(k(t-t_0))^2}
                               -\frac{3\sin k(t-t_0)}{(k(t-t_0))^3}\}].
\end{aligned}
\label{eq29}
\end{equation}
This is the fluctuations in energy density, from which the fluctuations in temperature can be obtained
by using the relation:  $\delta\rho/\bar{\rho}=4\delta T/T$ for $\rho\sim T^4$.  
We use Eqs.~\ref{eq17},\,\ref{f2} and \ref{eq29} to obtain the energy 
density fluctuation in terms of transport coefficients as:
\begin{equation}
\begin{aligned}
\Delta(\vec{k},t)=&e^{-(t-t_0)/\tau}[\Delta(\vec{k},t_0)
                              \frac{\sin k(t-t_0)}{k(t-t_0)}
                               +\frac{4}{k}\frac{\chi}{s}\frac{ T(\vec{k},t_0)}{\bar{T}}\{k^2 -ik_l\dot{U}_l(\vec{k},t_0)\}
                               \{\frac{\cos k(t-t_0)}{k(t-t_0)}- \frac{\sin k(t-t_0)}{k^2(t-t_0)^2}\}\\
                              &+\frac{40}{3}\frac{\eta}{s}\frac{\Theta(\vec{k},t_0)}{\bar{T}}
                              \{\frac{\sin k(t-t_0)}{k(t-t_0)}+\frac{3\cos k(t-t_0)}{k^2(t-t_0)^2}
                              -\frac{3\sin k(t-t_0)}{k^3(t-t_0)^3}\}].                        
\end{aligned}
\label{eq30}
\end{equation}
Eq.~\ref{eq30} provides the connection of the fluctuation in energy density in Fourier space with various 
transport coefficients {\it e.g.} thermal conductivity ($\chi$) and viscosity ($\eta$). 
It may be easily checked that the above solution satisfies the condition, 
$\Delta(\vec{k},t)\rightarrow \Delta(\vec{k},t_0)$ 
in the limit $t\rightarrow t_0$.  It is interesting to note that
the $k\sim 0$ mode (or large wave length mode) which is insensitive to spatial gradient 
is damped by the exponential time dependence only. 
$\Delta(k=0,t_0)$ represents the mode of the initial perturbation that takes the whole system (as in $k\rightarrow 0$, length scale of inhomogeneity $\lambda \rightarrow \infty$) slightly  away from its equilibrium value. However, the non-zero $k$ modes, in addition to the
exponential decay,
are damped out also due to spatial gradient  which is signalled by the presence of
terms involving shear viscosity and thermal conductivity in Eq.~\ref{eq30}.

The fluctuation in energy density in position space can be obtained by taking Fourier transformation of Eq.~\ref{eq30} as,
\begin{equation}
\frac{\delta\rho}{\bar{\rho}} (\vec{x},t)=\int \frac{d^3k}{(2\pi)^3} \Delta (\vec{k},t)\exp {(i\vec{k}.\vec{x})}
\label{eq31}
\end{equation}
If the initial ($t=t_0$) energy density fluctuation,  gradient of velocity, viscosity to entropy ratio and 
temperature of the system in equilibrium are  known then  Eqs.~\ref{eq30} and 
~\ref{eq31} can be used to get  fluctuations at any  time,  $t>t_0$. Angular
correlation function for these fluctuation has been estimated in the appendix.

\section{Evolution of fluctuation in a hydrodynamically expanding QGP background}
In the previous section we have considered the evolution of the fluctuations in a non-expanding
background. However, in a realistic scenario in RHIC-E the system expands due to  high 
internal pressure.  Therefore, in this section we include the effects of the expansion on the 
spatial anisotropy through the solutions of relativistic hydrodynamics. The fluid velocity 
and all the thermodynamic quantities become function of space time coordinates  to be determined
by the solution of the hydrodynamical equations. 

The space time variation of quantities such as energy density and flow velocity is
governed by relativistic hydrodynamics. Therefore, we solve the equation:
\begin{equation}
\partial_\mu \bar{T}^{\mu\nu}=0
\label{hydro}
\end{equation}
with the 
assumption that net baryon number density ($n_B$) is zero at the central rapidity region, hence
we need not consider the equation $\partial_\mu(n_Bu^\mu)=0$.
We also assume boost invariance~\cite{bjorken} along the longitudinal direction
and solve  the Eq.~\ref{hydro} numerically with equation of state $\bar{P}=\bar{\rho}/3$ 
for initial condition taken from optical Glauber
at the highest RHIC energy ($\sqrt{s_{NN}}=200$ GeV) for Au+Au collision. 
The hydrodynamic solutions~\cite{sushant} for flow velocity and temperature $(\bar{T}=[30\bar{\rho}/(g\pi^2)]^{1/4})$ 
have been used to study the space time evolution of the fluctuations in an expanding background.    

The evolution of  $\delta f (=f_0\psi)$ is governed by
the  equation, $p^{\mu}\partial_{\mu}f=(p\cdot u)C[f]$ which reduces to the following for an expanding
system  in the relaxation time approximation:
\begin{equation}
\left( \frac{\partial}{\partial t}+\frac{\vec{p}}{p^0}\cdot\frac{\partial}{\partial \vec{x} }
+\frac{(p^0u_0-\vec{p}\cdot\vec{u})}{p^0 \tau_R(x)}\right)\delta f(x,p)
=-\left( \frac{\partial}{\partial t}+
\frac{\vec{p}}{p^0}\cdot \frac{\partial}{\partial \vec{x} }\right)f_0(x,p)
\label{bterelax}
\end{equation}
The solution of Eq.~\ref{bterelax} is given by~\cite{hlevine}:
\begin{equation}
\delta f(x,p)=D(t,t_0)\left[ \delta f_{in}(p,\vec{x}-\frac{\vec{p}}{p^0}(t-t_0))+
\int_{t_0}^t B(\vec{x}-\frac{\vec{p}}{p^0}(t-t'),t') \,D(t_0,t') dt'\right]
\label{btesoln}
\end{equation}
where 
\begin{equation}
D(t_2,t_1)=\exp\left[-\int_{t_1}^{t_2} dt^\prime A(p,\,\vec{x}-\frac{\vec{p}}{p^0}(t'-t_0),\,t')\right]
\end{equation}
with
\begin{equation}
A(p,\vec{x},t)=\frac{u_0(x)-\vec{p}\cdot\vec{u}(x)/p_0}{\tau_R(x)}
\end{equation}
and
\begin{equation}
B(\vec{x},t)=-\left( \frac{\partial}{\partial t}+
\frac{\vec{p}}{p^0}\cdot \frac{\partial}{\partial \vec{x} }\right)f_0(x,p)
\end{equation}
For
\begin{equation}
f_0(x,p)=f_{eq}=\frac{1}{e^{\beta(x)(u^\mu p_\mu)}-1}
\end{equation}
The expression for $B$ reduces to:
\begin{equation}
B(\vec{x},t)=-f_{eq}(1+f_{eq})\frac{p^{\mu}}{p^{0}}\partial_{\mu}\left[\beta(x)u^\mu p_\mu\right]
\end{equation}
We took the relaxation time as $\tau_R^{-1}(x)=1.1\alpha_s T(x)$~\cite{thoma} (we have
taken constant value of $\alpha_s=0.2$ here), 
$\beta={1}/{T(x)}$,
$u^\mu(x)=(\gamma,\gamma\,\vec{v})$ is the four velocity of the fluid  and
$\gamma(x)=u^0(x)=(1-v(x)^2)^{-1/2}$.   The interaction of the expanding 
background with the fluctuation is affected through the relaxation time which depends 
on $T$ and the space-time variation of the temperature and 
velocity fields are determined by  the solution of the relativistic hydrodynamic equations.  
Eq.~\ref{btesoln} provides the space-time evolution of fluctuation in phase space distribution
for an expanding QGP background. This equation may be used to estimate various auto-correlations
and fluctuations  in thermodynamic quantities which can be measured experimentally. For example, the
fluctuation in temperature may be estimated by using Eqs. 36 and 15 which may be connected to fluctuations
in the transverse momentum (Ref. 30) measured experimentally.

To simulate initial spatial anisotropy with different geometry, we choose,
\begin{equation}
\delta f(p,\vec{x},t_0)=A_0\exp\left[-r(1+a_n \cos n\phi)\right]
\label{initialdist}
\end{equation}
We have taken $n= 2,3,4,5$ and $11$ to simulate different initial anisotropy. 
$A_0$ is set to unity for numerical results discussed below.   
From the solution in coordinate space one can get Fourier modes of fluctuations using Fourier 
Transformation which will give evolution of different Fourier modes of fluctuations. We take $a_n=0.3$ for $n=2,3,4,5$ and $11$.

\section{Results}
In this work we have used BTE to study the anisotropic fluctuations. 
The description of the evolution of anisotropy induced fluctuations within the ambit
of kinetic theory approach helps in getting better microscopic insight on the 
evolution. Moreover, kinetic theory approach has validity over
a wider range of phase space compared to hydrodynamical descriptions.
In the two subsections below we present results on the evolution  of fluctuations in a static 
and subsequently for a realistic scenario of expanding backgrounds respectively. 

\begin{figure}
\centerline{\includegraphics[height=60mm, width=60mm]{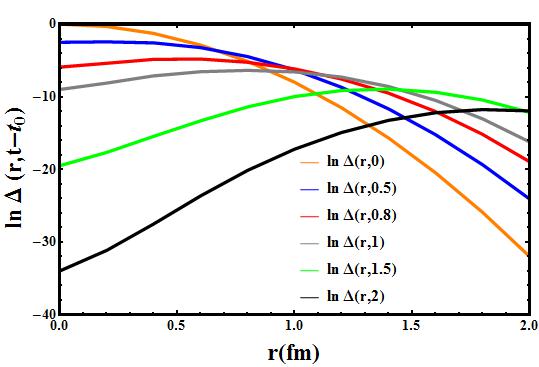}}
\caption{Evolution of the fluctuation in energy density with $r$ at different $t$ for a 
non-expanding QGP background.
}
\label{fig1a}
\end{figure}
\begin{figure}
\centerline{\includegraphics[height=60mm, width=60mm]{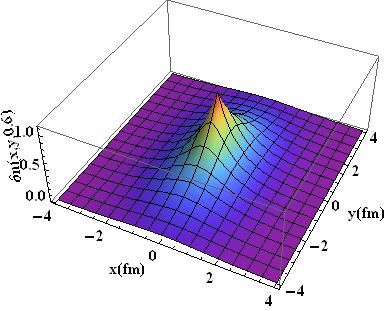}}
\centerline{\includegraphics[height=60mm, width=60mm]{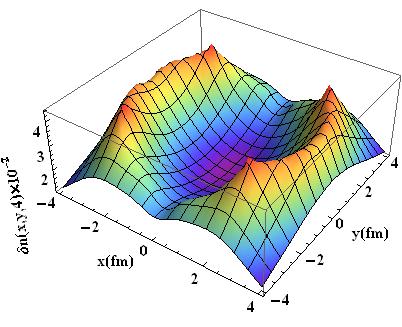}}
\caption{Evolution of the spatial anisotropy of the perturbation with initial elliptic geometry 
at time $\tau_0=0.6$ fm/c (upper panel). The lower panel shows the geometry after a time 4 fm/c 
has elapsed. Hydrodynamic expansion of the QGP background has been taken into account. 
Here the boundary of the background has an elliptic shape with the dimension of major and minor
axes are approximately $6$ fm and $4$ fm respectively.
The colours from red to violet represent highest to lowest values of the perturbations.
}
\label{fig1}
\end{figure}
\begin{figure}
\centerline{\includegraphics[height=60mm, width=60mm]{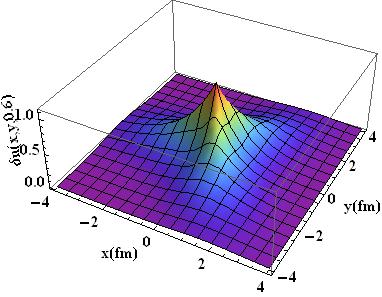}}
\centerline{\includegraphics[height=60mm, width=60mm]{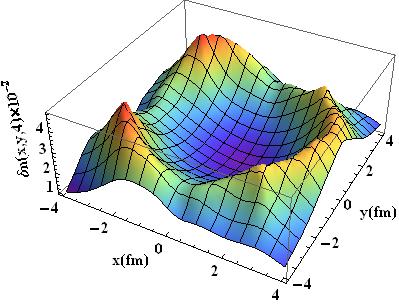}}
\caption{Same as Fig.~\ref{fig1} but the spatial anisotropy has a triangular geometry at 
the initial time $\tau_0=0.6$ fm/c (upper panel).  The lower panel shows the geometry after 
a time 4 fm/c  has elapsed.  
}
\label{fig2}
\end{figure}

\begin{figure}
\centerline{\includegraphics[height=60mm, width=60mm]{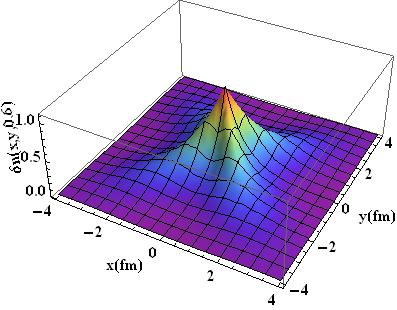}}
\centerline{\includegraphics[height=60mm, width=60mm]{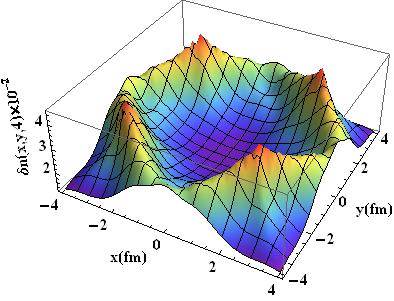}}
\caption{Same as Fig.~\ref{fig1} but the spatial anisotropy has a quadrangular geometry at the initial time $\tau_0=0.6$ fm/c (upper panel).
The lower panel shows the geometry after a time 4 fm/c has elapsed.  
}
\label{fig3}
\end{figure}

\begin{figure}
\centerline{\includegraphics[height=60mm, width=60mm]{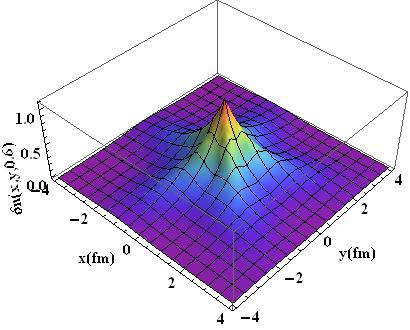}}
\centerline{\includegraphics[height=60mm, width=60mm]{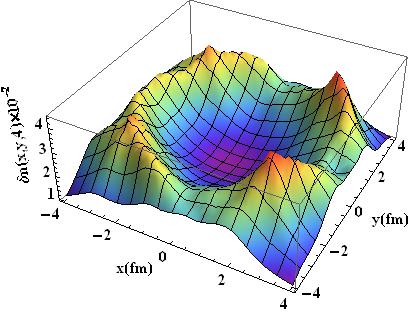}}
\caption{Same as Fig.~\ref{fig1} but the spatial anisotropy has a pentagonal geometry at the initial time $\tau_0=0.6$ fm/c (upper panel).
The lower panel shows the results after a time 4 fm/c has elapsed.  
}
\label{fig4}
\end{figure}
\begin{figure}
\centerline{\includegraphics[height=60mm, width=60mm]{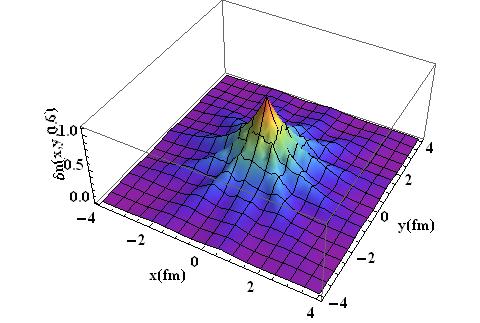}}
\centerline{\includegraphics[height=60mm, width=60mm]{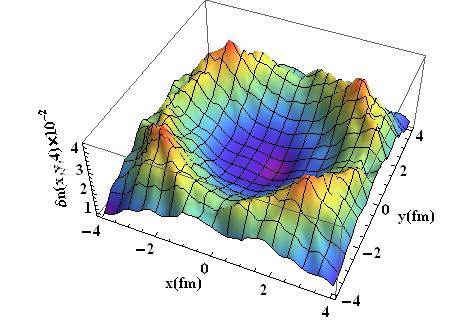}}
\caption{Same as Fig.~\ref{fig1} but the spatial anisotropy has a hendecagonal geometry
($n=11$) at the initial time $\tau_0=0.6$ fm/c (upper panel).
The lower panel shows the results after a time 4 fm/c has elapsed.  
}
\label{fig4a}
\end{figure}

\subsection{Evolution of fluctuation in energy density for a non-expanding system }
We display the spatial variation of the  fluctuation 
in Fig.~\ref{fig1a} at different times for a non-expanding background. 
We substitute Eq.~\ref{eq22a} in Eq.~\ref{fluc1} with $\Psi_{in}$ as a Gaussian in space
at the initial time and evaluate the evolution of the fluctuation. 
For the sake of illustration we take $T=400$ MeV and $\tau\sim 1$ fm/c.    
The results indicate a rapid dissipation and displacement of the peak of the fluctuations with 
increase in time. The displacement of the peak of the initial fluctuation given by Eq.~\ref{initialdist}  centered at 
$r=|\vec{x}|=0$ is governed by the factor,
$\vec{x}-\vec{p}(t-t_0)/p_0$ appearing in the solution for $\Psi$ (Eq.~\ref{eq22a}) and
the dissipation is controlled by the relaxation time, $\tau_R$ involves in the 
exponential factor  in the same equation. The dissipation will slow down in an expanding medium 
because the relaxation time will increase with decreasing temperature due to expansion.

\subsection{Evolution of fluctuation in an expanding QGP background}
In this section we would like to do some case study of how a given spatial anisotropy
characterized by some geometric shape will evolve with space and time in an expanding
QGP medium governed by relativistic hydrodynamics. This will give us some idea on the evolution of 
elliptical or triangular anisotropic perturbations created in the collisions.

We present the results now for a realistic scenario where the background QGP is expanding hydrodynamically
as described in section 3. 
Taking the value of the  temperature dependent relaxation time relevant for QCD plasma~\cite{thoma}  
the evolution of initial spatial anisotropies introduced through $\Psi_{in}$ (or $\delta f$) have been studied.
It is to be noted that due to expansion the temperature decreases and hence the relaxation time
increases which slows down the dissipation. 
Therefore, the dissipation of the perturbation gets slower with the expansion of the system. 
The effects of perturbation has better chance of survivability in the direction of lesser extent
because the expansion is faster along that direction due to larger pressure gradient. It implies 
that systems with same energy density the perturbations has larger chances to survive in systems with
smaller size. 
Then it is expected that the presence of perturbations will be dominant in relatively smaller size systems 
between events of same class.     
In Fig.~\ref{fig1} the evolution of the initial elliptic spatial anisotropy of the 
perturbation (upper panel), realized 
by taking $n=2$ in Eq.~\ref{initialdist} is depicted. The initial thermalization time is taken as
$\tau_0=0.6$ fm/c.  The evolution is studied up to $\tau=4$ fm/c. The red to 
violet colours used in the figures for distinct visibility, 
represent correspondingly the highest to lowest values of the perturbations. 
For $n=2$ the anisotropy has an elliptic shape having stronger gradient along $x$-axis resulting in 
faster expansion along $x$ compared to $y$ axis. Therefore, the propagation of the perturbation
generates a pattern similar to the one generated in water waves by the impact of a stick on the still
water surface. This kind of pattern is clearly observed in the lower panel of Fig.~\ref{fig1} such
type of fluctuation may be created by the propagation of jets through the QGP. 
It may be observed that the solution of BTE given in Eq.~\ref{btesoln} is subjected to two different kinds
of mechanism - (a) dissipation of the fluctuations and 
(b) hydrodynamic expansion of the background. The expansion velocity will be larger along $x$-axis
than along $y$-axis due to different pressure gradient imposed by the initial geometry 
of the fluctuation. This results in the splitting of the fluctuation as observed in the lower panel of Fig.~\ref{fig1}. 
By switching off the dissipation (appearing through the exponential term in $D(t,t_0)$) we have noticed 
that the fluctuation still splits in two parts but the peak of the fluctuation does not reduce significantly.    
It is also important to note that the two oppositely propagating perturbations are
correlated which may have interesting observation effects. 

Moreover, if the perturbation
is created  near the boundary of the system then the wave propagating  outward will  dissipate less
than the one moving inward.  
The results in Fig.~\ref{fig1} indicate a rapid dissipation of the peak. The peak has been reduced 
by more than $90\%$ at a time $\tau=4$ fm/c. 
The expansion of the QGP background is governed by the equation of state {\it i.e.} 
by the velocity sound in the QGP (the maximum displacement is determined 
by the sound horizon: 
{\it i.e.} the distance travelled by the sound wave: $\int_{\tau_0}^{\tau} d\tau c_s(\tau)d\tau$.
We have taken the sound velocity, $c_s=1/\sqrt{3}$, independent of $\tau$ for the expanding QGP background).
The displacement of the fluctuation (primarily $\delta f$) 
is regulated by the factor: $\vec{x}-\vec{p}(t-t_0)/p_0$ appearing in $\Psi$ (Eq.~\ref{eq22a}).  
Therefore, the net displacement is  determined by the combination of 
these two factors. 
The dissipation of the fluctuation is dictated by the relaxation time which is a function
of space-time coordinate through the relation: $\tau_R^{-1}\sim T(t,\vec{x})$. Therefore, the amplitude of
the displacement becomes a space time dependent quantity which is evident from the results displayed in 
Fig.~\ref{fig1}.  

The spatial anisotropic structure of the system formed in  RHIC-E can be understood 
with the help of Fourier analysis  in terms of its various coefficients. 
Work on the space-time evolution of the angular power spectrum for more realistic 
initial condition for the hydrodynamical solution derived from Glauber Monte-Carlo techniques 
is under progress~\cite{sarwar}. 

\begin{figure}
\centerline{\includegraphics[height=60mm, width=60mm]{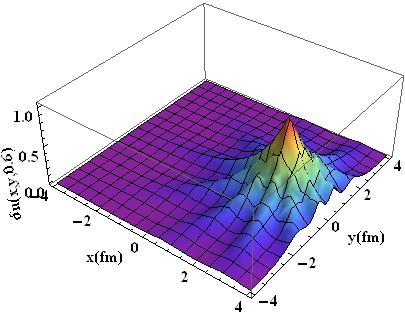}}
\centerline{\includegraphics[height=60mm, width=60mm]{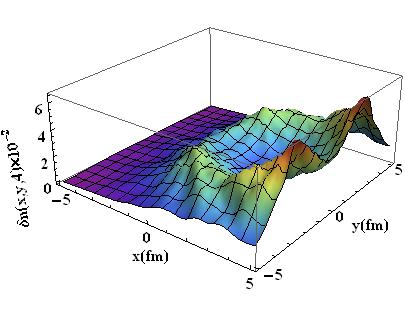}}
\caption{Same as Fig.~\ref{fig4a}  but the perturbation is given at 
a distance of 2.5 fm away from the origin along $x$-axis. 
The lower panel shows the results after a time 4 fm/c has elapsed.  
}
\label{fig6a}
\end{figure}
\begin{figure}
\centerline{\includegraphics[height=60mm, width=60mm]{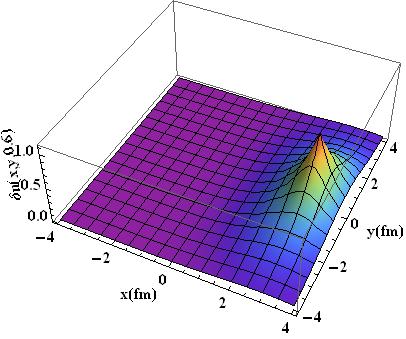}}
\centerline{\includegraphics[height=60mm, width=60mm]{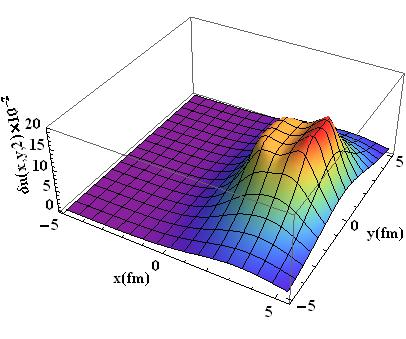}}
\centerline{\includegraphics[height=60mm, width=60mm]{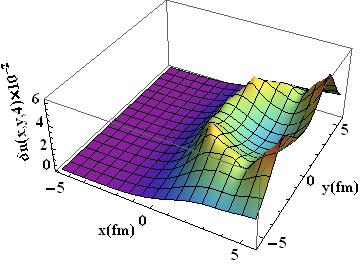}}
\caption{Same as Fig.~\ref{fig1} but the perturbation is given at 
a distance of 3 fm away from the origin along $x$-axis. 
The middle (lower) panel shows the results after a time 2 fm/c (4 fm/c) has elapsed.  
}
\label{fig7a}
\end{figure}
In Figs.~\ref{fig2} - \ref{fig4} the evolutions of the spatial anisotropic perturbations 
with different initial geometry like triangular, quadrangular and pentagonal for
$n=3,4$ and $5$ respectively have been depicted. We would like to see how these anisotropies
dissipate.  The perturbations introduce pressure gradient in the system.
The magnitude of the perturbation  gets reduced by the force arising due to pressure imbalance.  
It is observed that the spatial anisotropies of such perturbations dissipate fast.
The propagation of these anisotropic perturbations are affected primarily by the velocity of 
sound in the QGP background as well as by the velocity of
the fluctuation appearing in $\Psi$ as $\vec{x}-\vec{p}(t-t_0)/p_0$, and hence, on the 
thermal mass of the degrees of freedom that constitute the perturbation. The splitting
of the peaks are resulted from the expanding background with different magnitude
of velocities  due to different pressure gradient imposed by the initial geometry
of the fluctuation. 
The propagating waves for the perturbation take shape
analogous to water waves created on the calm surface if perturbed initially with
similar geometric shape.
Theoretical analysis of the angular power spectrum of the 
anisotropies arising from such perturbations in the evolving stage will shed light on possibility of selecting out the signatures of 
the early stage of the evolving matter.

In Fig.~\ref{fig4a} we display an initial perturbation (introduced at $r=0$) with smaller
angular dimension implemented through a hendecagonal ($n=11$) geometric shape
to  check whether such perturbations survive the evolution (upper panel).
The fate of the perturbation after space-time evolution is depicted
in the lower panel of Fig.~\ref{fig4a}. We observe that the perturbations of small 
angular size dissipate substantially. In fact, the perturbation with size corresponding to $n=5$  and $n=11$ look
similar at a time 4 fm/c after the initial time. We introduce the initial perturbation at distance 2.5 fm away from 
the origin along the positive $x$-axis (upper panel, Fig.~\ref{fig6a}). It is clear from the results displayed in Fig.~\ref{fig6a} (lower panel)
that the perturbation moving outward (away from  the centre) 
has suffered less dissipation compared to the one
propagating inward  and hence has a better chance to carry detectable signature. 

An elliptic perturbation is imparted near  the boundary  (Fig.~\ref{fig7a}, upper panel), 
3 fm away from the centre along the (positive) $x$-axis. The fate of the perturbation
after 2 fm/c and 4 fm/c are  shown in the middle and lower panels  of Fig~\ref{fig7a} respectively. 
It is interesting to note that the perturbation propagating away from the center  dissipates  less
and the one moving toward the centre of the background QGP decay fast.  Therefore, if  any perturbation
is created near the boundary the possibility of getting it detected is more. 

\begin{figure}
\centerline{\includegraphics[height=60mm, width=60mm]{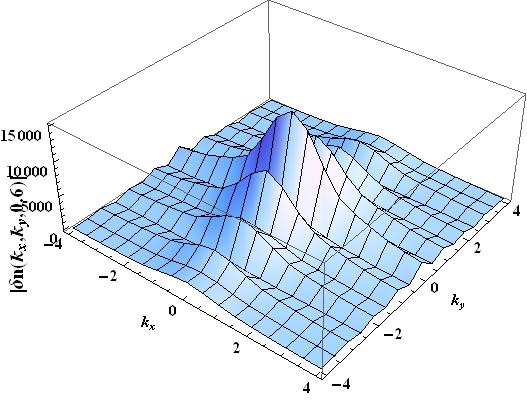}}
\centerline{\includegraphics[height=60mm, width=60mm]{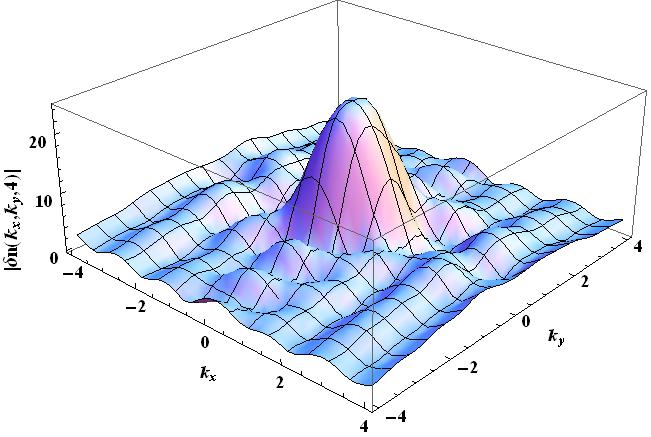}}
\caption{
(upper panel). The fluctuations in $k$-space at time $\tau=0.6$ fm/c (upper panel).
The lower panel shows the results after a time 4 fm/c has elapsed. The mixing of $k$-modes 
is visible in the lower panel.    
}
\label{fig5}
\end{figure}

\begin{figure}
\centerline{\includegraphics[height=60mm, width=60mm]{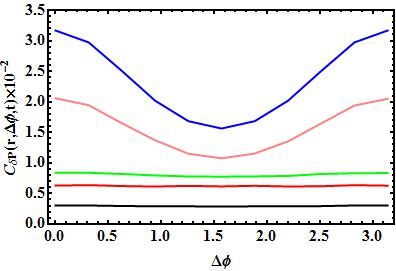}}
\caption{The angular auto-correlation of pressure is shown at $r=3$ fm as a function of $\Delta\phi$.  
The $\Delta\phi$ variation of $C_{\delta P}$ is displayed  at  different times  such as 0.6 fm/c (blue)
1 fm/c (orange), 2.5 fm/c (green), 3.5 fm/c (red)  and 4 fm/c (black line).
}
\label{fig6}
\end{figure}
The evolution of fluctuation, $\delta n(k_x,k_y,t)$ obtained by integrating $\delta\tilde{f}$ over 
$p$  in transverse $k$-space is depicted in Fig.~\ref{fig5} for
the initial shape at time 0.6 fm/c realized with  $n=2$ (Eq.\ref{initialdist}). We observe that the  pattern of the 
perturbation changes substantially from its initial distribution (upper panel)  
due to the mixing of various $k$-modes (lower panel) at a later time (4 fm/c).  The perturbation is 
propagating over a hydrodynamically  expanding background which makes all the variables like, 
temperature ($\bar{T}$), flow velocity ($v$), pressure ($\bar{P}$), energy density ($\epsilon$), etc
explicit functions of time and space. The interaction of the perturbation ($\delta f$) with the background
is incorporated through the relaxation time which is a function of temperature and 
hence space-time coordinates. Therefore, it is expected that various modes of the perturbations in
the Fourier space will get mixed during its propagation over the expanding background as clearly 
visible in Fig.~\ref{fig5} (lower panel). The peak of the fluctuation has reduced significantly
due  the exponential factor determined by the relaxation time.

\subsection{Correlation in pressure fluctuation and bulk viscosity}
The  auto-correlation function for fluctuation in pressure arising from perturbations is defined as: 
\begin{equation}
C_{\Delta P}(r,t,\Delta\phi)=\int d\phi\, \delta P(r,\phi,t)\,\delta P(r,\phi+\Delta\phi,t) 
\end{equation}
We evaluate $C_{\Delta P}$ at fixed $r(=3$ fm here) as a function of $t$ and $\Delta\phi$. 
The variation of $C_{\Delta P}$ with the angular separation $\Delta\phi$ is plotted in
Fig.~\ref{fig6} for perturbation with elliptic geometry ($n=2$) at different
times as indicated.
For $t=0.6$ fm/c the correlation function decreases with $\Delta\phi$ 
attains a dip at $\Delta\phi\sim \pi/2$ and again increases to produce a symmetric
behaviour about the dip. At $t=1$ fm/c the $\Delta\phi$ variation of $C_{\Delta P}$ is similar to
earlier time with an overall reduction in the magnitude.
At a later time, $t=2.5$ fm/c the $C_{\Delta P}$ evolves to a plateau.
This indicates that the power 
spectrum, $[\delta\tilde{P(k)}]^2$ is a Dirac delta function. 
It is also interesting to note that the $C_{\Delta P}$ at a given $r$ and $\Delta\phi$ 
decreases monotonically with time {\it i.e.} the correlation becomes weaker 
in real space as the perturbation reduces and the system approaches toward equilibrium  
with the progress of time. 
The evolution of correlation in the pressure fluctuation is crucial  for 
the study of flow harmonics in RHIC-E. 
The angular differential pressure will give rise to various non-zero 
flow coefficients like, elliptic, triangular and higher orders. 
The measured anisotropy can be extrapolated backward in time through theoretical model to characterize the 
early state of the matter formed in RHIC-E. 

\begin{figure}
	\centerline{\includegraphics[height=60mm, width=60mm]{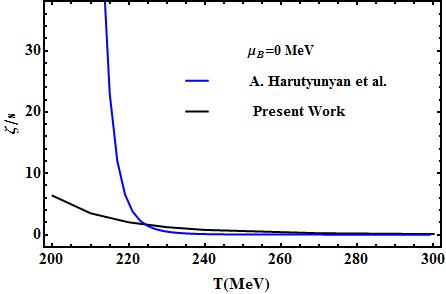}}
	\caption{The variation of bulk viscosity to entropy ratio ($\zeta/s$) as a function
of temperature. 
	}
	\label{fig8}
\end{figure}

The bulk viscosity of matter created in RHIC-E is a field of high contemporary interest~\cite{QGPbulk}.
We use the current formalism to estimate the bulk viscous coefficient ($\zeta$). 
The fluctuations in thermodynamic quantities can be used to estimate various transport coefficients. 
For example, the fluctuations in pressure ($\delta P<< \bar{P}$) determined by the $\delta f$ (Eq.~\ref{eq22a}) can be employed to 
calculate the bulk viscous coefficient (several methods have been employed in the literature
to estimate bulk viscosity of QGP some of these are discussed in~\cite{QGPbulk})  of the quarks with thermal mass~\cite{lebellac} by using  
Green-Kubo relation~\cite{Gkubo}  in the domain of linear response. 
The bulk viscosity ($\zeta$) is related to the correlation of time dependent pressure fluctuation as follows~\cite{NEx}: 
\begin{equation}
\zeta=\frac{V}{T}\int_0^\infty dt \langle \delta P(t) \delta P(0)\rangle
\end{equation}
We estimate the $\zeta$ by using this relation and compare the bulk 
viscosity to entropy density ($s$) ratio as a function of
temperature to the results obtained in Ref.~\cite{Gkb2f} 
in the strong coupling limit with two flavour NJL model (Fig~\ref{fig8}). 
We observe that the behaviour of $\zeta/s$ in the high 
$T (>225$ MeV) regime is similar to that obtained in Ref.~\cite{Gkb2f}. 
This is reasonable because the relaxation time used in the present work
has been estimated for weakly coupled QGP~\cite{thoma} 
which may be realized at the high $T$ regime. The $\zeta/s$ calculated in~\cite{Gkb2f}
rises very fast with lowering of $T$  (for $T<225$ MeV) due to multi-loop contributions,
inclusion of such contributions is beyond the scope of the present work. 
However, it has been verified that the $\zeta/s$  obtained here is similar 
to the $\zeta/s$ reported in Ref.~\cite{Gkb2f} with single loop contribution which may
be a good approximation for weakly coupled system. 

\subsection{Temperature fluctuation}
The solution of the BTE, $\delta f$ can be used to estimate the 
fluctuation in useful thermodynamic quantities in QGP. 
For example, the fluctuation
in temperature, $\Delta T$ in different azimuthal bins ($\Delta\phi$) can be
calculated as follows. 
Since the average transverse momentum ($\langle p_T \rangle$) is directly proportional to the 
temperature of the QGP 
the fluctuation in temperature in a bin $\Delta\phi$  is given  by 
the  relation,
\begin{equation}
\Delta T  \sim\int_{\phi_1}^{\phi_2}d\phi\int_{0}^{\infty}p_Tdp_T\int d^3x 
\delta f
\end{equation}
Therefore, the $\Delta\phi$ variation of temperature fluctuation in 
little bang {\it i.e.} for the system formed in RHIC-E  can be estimated
analogous to the temperature fluctuation in the universe in the recombination 
era. These issues along with the power spectrum similar to the CMBR 
will be addressed in Ref.~\cite{sarwar}.  

\section{Summary and Discussion}
The evolution of fluctuations have been studied  
in Refs~\cite{pstaig,UW,kapusta,ripples,RJ,kovtun,luzum} using 
relativistic hydrodynamical model.  In contrast we use a more
microscopic approach to investigate the evolution of fluctuations
within the framework of BTE in a relativistically expanding QGP background. 
The background of the spatial fluctuations
has been assumed as a thermalized expanding QGP. The expansion
of the background has been dealt with the (2+1) dimensional
relativistic hydrodynamical model. 
The evolution of initial spatial
anisotropic perturbations with different geometry have been studied 
and analytical results have been obtained. 
It is found that the perturbations dissipate during its propagation, however, 
the creation of such  anisotropic perturbations near the boundary of the plasma may lead to 
detectable effects.  Theoretical analysis of 
these anisotropies  will help in understanding the early stage of the matter. 
We have provided an explicit relations between the fluctuations and transport coefficients.
The mixing of various $k$-modes of the perturbations during the course of evolution has been demonstrated.
The evolution of correlations of  perturbation in pressure has been studied and shown that the
correlation between two points in real space reaches a plateau at later time. 
We have used the calculated correlation in pressure fluctuation to estimate
the bulk viscous coefficient~\cite{okumura}.
The results presented here may be used 
to estimate correlations of multiplicity fluctuation at the freeze-out surface which
may be measured in experiments to extract transport coefficients like shear viscosity
as shown in Ref.~\cite{kapusta}. 
The power spectrum of temperature fluctuations (Eq.~\ref{delT})
can be estimated and compared with the power spectrum of
transverse momentum fluctuation of thermal hadrons measured experimentally.
A detail analysis of the power spectrum with hydrodynamically
expanding background is underway~\cite{sarwar}, the pattern of which looks
similar to the power spectrum of angular momentum reported in Ref.~\cite{powerpt}.
This  will provide insight into the fate of the anisotropies created initially in the system produced
in RHIC-E.  A comment on the present study may be made here. Although in the present work 
we have discussed the evolution spatial anisotropy in an expanding QGP background, the formalism 
discussed is relevant for studying space-time evolution of fluctuations in any relativistically
expanding background.  
\\

{\noindent{\bf Acknowledgement:} G. S. acknowledges the support from Department of Atomic Energy, Govt. of India for
this work. We would like to thank Mr Sushant Kumar Singh for providing inputs from relativistic hydrodynamical code.}

\section{Appendix: Correlations}
In this appendix, we evaluate the correlation, $\langle\Delta(\hat{n_1})\Delta(\hat{n_2})\rangle$.
The fluctuations, $\Delta(\vec{x},\hat{n},t)$ can be written as:
\begin{equation}
                 \Delta(\vec{x},\hat{n},t)=\sum _{l,m} a_{lm}(\vec{x},t) Y_{lm}(\hat{n}),\\
\label{eq32}
\end{equation}
with   $a_{lm} = (-i)^l4\pi \int d^3 \hat{n}\, Y^*_{lm}(\hat{n})\Delta(\vec{x},\hat{n},t)$ 
          and $\langle a_{lm}.a^*_{l'm'}\rangle =C_l(\vec{x},t)\delta_{ll'}\delta_{mm'}$.
Using Eq.~\ref{eq32} we can find the correlations in the following way,
\begin{equation}
\begin{aligned}
                \langle \Delta(\vec{x},\hat{n}_1,t)\Delta(\vec{x},\hat{n}_2,t)\rangle&=\sum_{l,m,l',m'}\langle a_{lm}.a^*_{l'm'}\rangle  Y_{lm}(\hat{n}_1) Y^*_{l'm'}(\hat{n}_2),\\
                &=\sum_{l,m,l',m'}C_l(\vec{x},t)\delta_{ll'}\delta_{mm'}  Y_{lm}(\hat{n}_1) Y^*_{l'm'}(\hat{n}_2),\\
                &=\sum_{l,m}C_l(\vec{x},t)  Y_{lm}(\hat{n}_1) Y^*_{lm}(\hat{n}_2),\\
\end{aligned}
\label{eq33}
\end{equation}

which leads to $$\langle \Delta(\vec{x},\hat{n}_1,t)\Delta(\vec{x},\hat{n}_2,t)\rangle =\frac{1}{4\pi}\sum_l(2l+1)C_l(\vec{x},t)P_l(\hat{n}_1\cdot\hat{n}_2),$$
since, $P_l(\hat{n}_1\cdot\hat{n}_2)=\frac{4\pi}{2l+1}\sum_{m=-l}^{l}  Y_{lm}(\hat{n}_1) Y^*_{lm}(\hat{n}_2)$.
Eq.~\ref{eq33} defines the correlations of fluctuations observed from two different directions in terms of co-efficient $C_l$s. $C_l$'s
are the angular power spectrum which contains the information of the anisotropies. Work is under progress to estimate these coefficients
by solving the hydrodynamical equations with the initial conditions taken from Glauber Monte-Carlo method.  
Similarly, one can define these co-efficients corresponding to $k$-space presentation of fluctuations. 
Time evolution of these co-efficients can be obtained from evolution of energy density fluctuation, 
$\Delta(\vec{k},\hat{n},t)$ given by Eq.~\ref{eq30}.
Therefore, we have
\begin{equation}
\begin{aligned}
 \langle \Delta(\vec{x},\hat{n}_1,t)\Delta(\vec{x},\hat{n}_2,t)\rangle&=\int \frac{d^3k}{(2\pi)^3} \frac{d^3k'}{(2\pi)^3} \ e^{i(\vec{k}-\vec{k'})\cdot \vec{x}}\ \langle \Delta (\vec{k},\hat{n}_1,t)\Delta (\vec{k'},\hat{n}_2,t)\rangle 
\end{aligned}
\label{eq34}
\end{equation}
Now,
\begin{equation}
\begin{aligned}
\langle \Delta(\vec{k},\hat{n}_1,t)\Delta(\vec{k'},\hat{n}_2,t)\rangle=&\langle \Delta(\vec{x},\hat{n}_1,t_0)\Delta(\vec{x},\hat{n}_2,t_0)\rangle L(\vec{k},t_0;\vec{k'},t_0) \\
&+\langle \Theta(\vec{x},\hat{n}_1,t_0)\Theta(\vec{x},\hat{n}_2,t_0)\rangle M(\vec{k},t_0;\vec{k'},t_0)\\
&+\langle \Delta(\vec{x},\hat{n}_1,t_0)\Theta(\vec{x},\hat{n}_2,t_0)\rangle  N(\vec{k},t_0;\vec{k'},t_0),
\end{aligned}
\label{eq35}
\end{equation}
where, 
\begin{equation}
\begin{aligned}
L(\vec{k},t,\vec{k'},t_0)=&e^{-\frac{2(t-t_0)}{\tau}}
                               \{ \frac{\sin k(t-t_0)}{k(t-t_0)}\}\{ \frac{\sin k'(t-t_0)}{k'(t-t_0)}\},\\
M(\vec{k},t,\vec{k'},t_0)=&e^{-\frac{2(t-t_0)}{\tau}}
                                               \left(\frac{40}{3}\frac{\eta}{s\bar{T}}\right)^2
                                              \{  \frac{\sin k(t-t_0)}{k(t-t_0)}+\frac{3\cos k(t-t_0)}{(k(t-t_0))^2} -\frac{3\sin k(t-t_0)}{(k(t-t_0))^3}\}\\
                                              &\times \{  \frac{\sin k'(t-t_0)}{k'(t-t_0)}+\frac{3\cos k'(t-t_0)}{(k'(t-t_0))^2} -\frac{3\sin k'(t-t_0)}{(k'(t-t_0))^3}\}\\
N(\vec{k},t,\vec{k'},t_0)=&e^{-\frac{2(t-t_0)}{\tau}}
                                              \frac{40}{3}(\frac{\eta}{s\bar{T}})
                                               \{ \frac{\sin k(t-t_0)}{k(t-t_0)}\}\\
                                              &\times \{  \frac{\sin k'(t-t_0)}{k'(t-t_0)}+\frac{3\cos k'(t-t_0)}{(k'(t-t_0))^2} -\frac{3\sin k'(t-t_0)}{(k'(t-t_0))^3}\}
\end{aligned}
\label{eq36}
\end{equation}
For two functions $\Delta$ and $\Theta$, defining the correlation as:
$\langle \Delta(\vec{k},\hat{n}_1,t)\Theta(\vec{k'},\hat{n}_2,t)\rangle=(2\pi)^3\delta(\vec{k}-\vec{k'})
\delta_{\Delta\Theta}\langle \Delta({k},\hat{n}_1,t)\Theta({k},\hat{n}_2,t)\rangle$ 
and $\Delta(\vec{k},\hat{n},t)=\sum _{l,m} a^{\Delta}_{lm}(\vec{k},t) Y_{lm}(\hat{n})$,
$\langle{ a^{\Delta}}_{lm}.{a^*}^{\Theta}_{l'm'}\rangle =C^{\Delta\Theta}_l(\vec{k},t)\delta_{ll'}\delta_{mm'}$, we get

\begin{equation}
\begin{aligned}
\langle \Delta(\vec{k},\hat{n}_1,t)\Delta(\vec{k'},\hat{n}_2,t)\rangle&=(2\pi)^3\delta(\vec{k}-\vec{k'})\sum_l\frac{2l+1}{4\pi}P_l(\hat{n}_1\cdot \hat{n}_2)\\
&\times \{C^{\Delta\Delta}_l({k},t_0)L(k,t,k,t_0)+C^{\Theta\Theta}_l({k},t_0)M(k,t,k,t_0)\}.\\
\end{aligned}
\label{eq37}
\end{equation}
\begin{equation}
\begin{aligned}
\text{Using}\hspace{.1in}\langle \Delta(\vec{k},\hat{n}_1,t)\Delta(\vec{k'},\hat{n}_2,t)\rangle&=(2\pi)^3\delta(\vec{k}-\vec{k'})\sum_l\frac{2l+1}{4\pi}C^{\Delta\Delta}_l({k},t)P_l(\hat{n}_1\cdot \hat{n}_2),\text{in Eq.~\ref{eq37}}\ \ \ 
\text{we get,}\nonumber
\end{aligned}
\end{equation}
\begin{equation}
 C^{\Delta\Delta}_l(\vec{k},t)=C^{\Delta\Delta}_l({k},t_0)L(k,t,k,t_0)+C^{\Theta\Theta}_l({k},t_0)M(k,t,k,t_0).
\label{eq38}
\end{equation}
Using Eq.~\ref{eq33},~\ref{eq38} in Eq ~\ref{eq34}, we get
\begin{equation}
\begin{aligned}
\ C^{\Delta\Delta}_l(\vec{x},t)&=\int d^3k\{  C^{\Delta\Delta}_l({k},t_0)L(k,t,k,t_0)+C^{\Theta\Theta}_l({k},t_0)M(k,t,k,t_0)\}
\end{aligned}
\label{eq39}
\end{equation}

This provides the correlations at different angular scales at time $t$ for a given correlations at initial time, $t_0$ ($t>t_0$). 

\end{document}